\documentclass[12pt,cite,a4paper]{article}

\usepackage{amsmath}
\usepackage{amssymb}
\usepackage[dvips]{graphicx}
\begin{document}

\begin{titlepage}
\begin{flushright}
IFUP--TH 2005/15 \\
\end{flushright}
~

\vskip .8truecm
\begin{center}
\Large\bf Quantum Liouville theory on the pseudosphere\\
 with heavy charges \footnote{This work is
supported in part
  by M.I.U.R.}
\end{center}

\vskip 1truecm
\begin{center}
{Pietro Menotti} \\
{\small\it Dipartimento di Fisica dell'Universit{\`a}, Pisa 56100,
Italy and}\\
{\small\it INFN, Sezione di Pisa}\\
{\small\it e-mail: menotti@df.unipi.it}\\
\end{center}
\vskip .8truecm
\begin{center}
{Erik Tonni} \\
{\small\it Scuola Normale Superiore, Pisa 56100, Italy and}\\
{\small\it INFN, Sezione di Pisa}\\
{\small\it e-mail: e.tonni@sns.it}\\
\end{center}

\vskip 2truecm

\begin{abstract}

We develop a perturbative expansion of quantum Liouville theory on
the pseudosphere around the background generated by heavy charges.
Explicit results are presented for the one and two point functions
corresponding to the summation of infinite classes of standard
perturbative graphs. The results are compared to the one point
function and to a special case of the two point function derived
by Zamolodchikov and Zamolodchikov in the bootstrap approach,
finding complete agreement. A partial summation of the conformal
block is also obtained.

\end{abstract}

\vskip 1truecm

\end{titlepage}

Much interest has been devoted to the exact solutions of the
conformal bootstrap equations for the Liouville theory on the
pseudosphere \cite{ZZpseudosphere}, on the finite disk with
conformally invariant boundary conditions \cite{FZZ} and on the
sphere \cite{ZZsphere, Dorn Otto, Teschner:
sphere}. The first two solutions have given rise to the ZZ and FZZT branes.\\
In this letter, we present a technique to treat the Liouville
field theory on the pseudosphere, which allows to find an
expansion in the coupling constant $b$ of the $N$ point functions
in presence of ``heavy charges'', according to the terminology
introduced in \cite{ZZsphere}. This means that we consider the
vertex operator $V_\alpha(z) = e^{2\alpha \phi(z)}$ with
$\alpha=\eta/b$ and $\eta$ fixed in the semiclassical limit $b
\rightarrow 0$. \\
For the one point function, this analysis goes well beyond the
previous perturbative expansion performed in \cite{ZZpseudosphere,
MTtetrahedron, MTgeometric} where $\alpha$ has been taken small;
indeed, our result corresponds to the summation of an infinite
class of perturbative graphs. Thus, we obtain a strong check of
the ZZ bootstrap formula for the one point function
\cite{ZZpseudosphere}, which includes all the
previous perturbative checks.\\
We apply the same technique to compute the two point function with
two heavy charges $\eta$ and $\varepsilon$, to the first order in
$\varepsilon$, getting a closed expression of this correlator to
the orders $O(b^{-2})$ and $O(b^0)$ included, but exact in $\eta$
and in the $SU(1,1)$ invariant distance $\omega$ between the
sources. According to an argument given by ZZ, such expression
provides an expansion of a conformal block up to $O(b^0)$ and to
the first order in $\varepsilon$, but to all orders in $\eta$ and
$\omega$. With more work, this technique can be extended to higher
orders in $b^2$ and to more
complicated correlation functions.\\

\noindent We start from the Liouville action on the pseudosphere
in presence of $N$ sources characterized by heavy charges
$\eta_1,\dots,\eta_N$, given in \cite{MTgeometric}. Decomposing
the Liouville field $\phi$ as follows
\begin{equation}
\phi
\,=\,\frac{1}{2b}\,\big(\,\varphi_{\scriptscriptstyle\hspace{-.05cm}B}
+ 2b\,\chi\,\big)
\end{equation}
the Liouville action separates into a classical part, depending
only on the background field
$\varphi_{\scriptscriptstyle\hspace{-.05cm}B}$, and a quantum
action for the quantum field $\chi$
\begin{equation}\label{action}
S_{\Delta,\,N}[ \,\phi\,]  \,=\,S_{cl}[
\,\varphi_{\scriptscriptstyle\hspace{-.05cm}B}\,] +
S_{q}[\,\varphi_{\scriptscriptstyle\hspace{-.05cm}B},\,\chi\,]\;.
\end{equation}
Adopting the unit disk representation $\Delta = \{z
\in\mathbb{C}\,; |z| < 1\}$ for the pseudosphere, these actions
read
\begin{eqnarray}\label{action classical phiB}
\hspace{-.2cm} S_{cl}[
\,\varphi_{\scriptscriptstyle\hspace{-.05cm}B}\,] & =
&\frac{1}{b^2}
 \lim_{\begin{array}{l}
\vspace{-.9cm}~\\
\hspace{0cm} \vspace{-.4cm} \scriptscriptstyle \varepsilon \rightarrow 0 \\
\hspace{.1cm} \scriptscriptstyle \!\!r \rightarrow 1\end{array}}\,
\Bigg\{ \;\int_{\Delta_{r,\varepsilon}} \left[ \,\frac{1}{4\pi}
\,\partial_z \varphi_{\scriptscriptstyle\hspace{-.05cm}B}
\,\partial_{\bar{z}}\varphi_{\scriptscriptstyle\hspace{-.05cm}B}+\mu
b^2
e^{\,\varphi_{\hspace{-.02cm} \scriptscriptstyle{B}}}\,\right]\, d^2 z  \, 
\nonumber \\
&  &\hspace{2.3cm} -\,\frac{1}{4\pi i}\; \oint_{\partial\Delta_r} \varphi_B
\left( \, \frac{\bar{z}}{1-z\bar{z}} \,d z-
       \,\frac{z}{1-z\bar{z}} \,d \bar{z} \right)+ f_{cl}(r,\mu b^2)\\
 \rule{0pt}{1cm} &  & \hspace{2.3cm}
-\,\frac{1}{4\pi i}\;\sum_{n=1} ^N \eta_n\oint_{\partial\gamma_n} \varphi_B
\left( \, \frac{d z}{z-z_n}- \frac{d \bar{z}}{\bar{z}-\bar{z}_n}\,
\right)- \sum_{n=1} ^N \eta_n^2 \log \varepsilon_n^2 \;\; \Bigg\}
\nonumber
\end{eqnarray}
and
\begin{eqnarray}\label{quantum action final}
\vspace{-1cm}
 S_{q}[\,\varphi_{\scriptscriptstyle\hspace{-.05cm}B}\,, \chi\,] & = &
 \lim_{\scriptstyle r\rightarrow 1}
\, \Bigg\{\; \int_{\Delta_{r}} \left[ \,\frac{1}{\pi} \,\partial_z
\chi \,\partial_{\bar{z}}\chi+\mu\, e^{\,\varphi_{\hspace{-.02cm}
\scriptscriptstyle{B}}}\,\big(\,e^{2b\,\chi}-1-2b\,\chi\,\big)\,\right]\,
d^2 z\\
 &  &\hspace{4.8cm} -\,\frac{b}{2\pi i}\;
\oint_{\partial\Delta_r}
  \chi\,
\left( \, \frac{\bar{z}}{1-z\bar{z}} \,d z\,- \frac{z}{1-z\bar{z}}
\,d \bar{z} \right)
       \;\,\Bigg\} \nonumber
\end{eqnarray}
where the function $f_{cl}(r,\,\mu b^2)$ is a subtraction term
independent of the charges. The coupling constant $b$ is related
to the parameter $Q$ occurring in the central charge $c=1+6Q^2$ by
$Q=1/b+b$ \cite{CT}. Moreover, the classical field
$\varphi_{\hspace{-.02cm} \scriptscriptstyle{B}}$ obeys the
following boundary conditions
\begin{eqnarray}
\label{varphiB bc infinity}
\varphi_{\scriptscriptstyle\hspace{-.05cm}B}(z) & = & -\, \log\,
(\,1-z\bar{z}\,)^2 + f(\mu b^2) + O\big((1-z\bar{z})^2\big)
\hspace{.9cm}\textrm{when}\hspace{.4cm} |z|\rightarrow 1 \\
\label{varphiB bc sources}
\varphi_{\scriptscriptstyle\hspace{-.05cm}B}(z)  & = & -\;2\eta_n
\log \,|\,z-z_n\,|^2\,+O(1)
\hspace{3.4cm}\textrm{when}\hspace{.4cm}\,z\,\rightarrow z_n
\end{eqnarray}
where $f(\mu b^2)$ is a function depending only on the product
$\mu b^2$. The integration domains are
$\Delta_{\,r,\,\varepsilon}=\Delta_{\,r}\hspace{-.07cm} \setminus
\bigcup_{n=1}^N \gamma_{n}$ with $\Delta_{\,r}=\{\,|z| < r < 1
\,\} \subset \Delta$ and $\gamma_{n}=\{|z-z_n|<\varepsilon_n\}$.
Because of the boundary behavior of the Green function, that will
be computed in the following, the last integral in the quantum
action (\ref{quantum action final}) does not contribute.\\
The vanishing of the first variation  of $S_{cl}[
\,\varphi_{\scriptscriptstyle\hspace{-.05cm}B}\,]$ with respect to
the field $\varphi_{\scriptscriptstyle\hspace{-.05cm}B}$
satisfying (\ref{varphiB bc infinity}) and (\ref{varphiB bc
sources}) gives the Liouville equation in presence of $N$ sources
\begin{equation}\label{liouville eq with sources varphi}
    -\,\partial_{z}\partial_{\bar{z}}\,\varphi_{\scriptscriptstyle
\hspace{-.05cm}B}\,
    +
    2\pi b^2 \mu\,e^{\,\varphi_{\hspace{-.02cm}
    \scriptscriptstyle{B}}} \,=\,
    2\pi \sum_{n=1}^N \eta_n
    \,\delta^2(z-z_n)\;.
\end{equation}
At semiclassical level, we have
\begin{equation}\label{semiclassic N point}
\left\langle \, V_{\alpha_1}(z_1)\dots V_{\alpha_N}(z_N)
\,\right\rangle_{sc}\,=\,
\frac{e^{-S_{cl}(\eta_1,\,z_1;\dots;\,\eta_N,\,z_N)}}{e^{-S_{cl}(0)}}
\end{equation}
where $S_{cl}(\eta_1,z_1;\dots;\eta_N,z_N)$ is the classical
action $S_{cl}[ \,\varphi_{\scriptscriptstyle\hspace{-.05cm}B}\,]$
computed on the solution
$\varphi_{\scriptscriptstyle\hspace{-.05cm}B}$ of the Liouville
equation with sources. Now, since under a $SU(1,1)$ transformation
the classical background field changes as follows
\begin{equation}\label{varphiB transformations}
\varphi_{\scriptscriptstyle\hspace{-.05cm}B}(z)
\hspace{.5cm}\longrightarrow  \hspace{.5cm}
\tilde{\varphi}_{\scriptscriptstyle\hspace{-.05cm}B}(w)
\;=\;\varphi_{\scriptscriptstyle\hspace{-.05cm}B}(z)\,-\, \log
\left| \frac{d w}{d z}\right|^{2}\;
\end{equation}
one can see that the transformation law of $S_{cl}[
\,\varphi_{\scriptscriptstyle\hspace{-.05cm}B}\,]$ assigns to the
vertex operator $V_\alpha(z)$ the semiclassical dimensions
$\alpha\,(1/b-\alpha)=\eta\,(1-\eta)/b^2$ \cite{MTgeometric, Takhtajan}.\\

\noindent For the one point function, we have a single heavy
charge $\eta_1=\eta$, which can be placed in $z_1=0$, and the
explicit solution of the Liouville equation is
$\varphi_{\scriptscriptstyle\hspace{-.05cm}B}=\varphi_{cl}$, given
by \cite{Seiberg: Notes}
\begin{equation}\label{phiclassic}
    e^{\varphi_{cl}}\,=\,\frac{1}{\rule{0pt}{.4cm}\pi\mu b^2}\;
    \frac{(1-2\eta)^2}{\big[\,\rule{0pt}{.4cm}(z\bar{z}
\hspace{.03cm})^{\eta}-(z\bar{z}\hspace{.03cm})^{1-\eta}\,\big]^2}\;.
\end{equation}
Local finiteness of the area around the source in the metric
$e^{\varphi_{cl}} \,d^2z$ imposes $\eta<1/2$ \cite{Seiberg: Notes, Picard}.\\
The classical action (\ref{action classical phiB}) computed on
this background gives the semiclassical one point function
\begin{equation}\label{one point classical term}
\left\langle \, V_{\eta/b}(0) \,\right\rangle_{sc} \,=\,
\exp\left\{-\,\frac{1}{b^2}\;\Big(\,\eta\,\log\left[\,\pi
        b^2\mu\,\right]+2\eta+(1-2\eta)\,\log(1-2\eta)\,\Big)\,\right\}\;.
\end{equation}
To go beyond this approximation, we need to find the Green
function on the background field given by (\ref{phiclassic}) and,
to do this, we employ the method developed in \cite{MV}. Thus, we
compute the classical background in presence of the charge $\eta$
in $z_1=0$ and of another charge $\varepsilon$ in $z_2=t \in
\Delta$ and real by applying first order perturbation theory
in $\varepsilon$ to the fuchsian equation associated to the
Liouville equation, i.e.
\begin{equation}\label{fuchsian eq}
    \frac{d^{\,2} Y_j}{dz^2}\,+ Q(z)\,Y_j\,=\,0
    \hspace{2cmj\,=\,1,\,2}
\end{equation}
where $Q(z)=b^2 T(z)$, being $T(z)$ the holomorphic component of
the classical energy momentum tensor of the classical field
$\varphi_{\scriptscriptstyle\hspace{-.05cm}B}$. To first order
perturbation theory in $\varepsilon$, one writes
$Y_j=y_j+\varepsilon\,\delta y_j$ and $Q=Q_0+\varepsilon\,q$,
where the unperturbed quantities are $Q_0(z)=\eta(1-\eta)/z^2$,
$y_1(z)=z^\eta$ and $y_2(z)=z^{1-\eta}$.\\
We impose now the Cardy condition \cite{Cardy} and the regularity
condition at infinity on the classical energy momentum tensor. One
can express them more easily in the upper half plane
$\mathbb{H}=\{\,\xi\in\mathbb{C}\,;\, \textrm{Im}(\xi)>0\,\}$
representation (related to the unit disk representation through
the Cayley transformation $z=(\xi-i)/(\xi+i)$), where they read
$\widetilde{Q}(\xi)\,=\,\overline{\rule{0pt}{.45cm}\widetilde{Q}}(\xi)$
and $\xi^4\,\widetilde{Q}(\xi) \,\sim\,
O(1)$ when $\xi\rightarrow\infty$, respectively.\\
In the $\mathbb{H}$ representation, we have that
$\widetilde{Q}_0(\xi)=4\eta(\eta-1)/(\xi^2+1)^2$ and
\begin{equation}\label{q on H}
\tilde{q}(\xi)\,=\, \frac{1}{\rule{0pt}{.4cm}(\,\xi-i\,\tau\,)^2}
\,+\,\frac{1}{\rule{0pt}{.4cm}(\,\xi+i\,\tau\,)^2}
\,+\,\frac{\beta_{i}}{\rule{0pt}{.4cm}2\,(\,\xi-i\,)}
\,+\,\frac{\bar{\beta}_{i}}{\rule{0pt}{.4cm}2\,(\,\xi+i\,)}
\,+\,\frac{\beta_{i\tau}}{\rule{0pt}{.4cm}2\,(\,\xi-i\,\tau\,)}
\,+\,\frac{\bar{\beta}_{i\tau}}{\rule{0pt}{.4cm}2\,(\,\xi+i\,\tau\,)}
\phantom{***}
\end{equation}
where $\beta_i$ and $\beta_{i\tau}$ are the Poincar\'e accessory
parameters and $i\tau=i(1+t)/(1-t)$ is the image of $z_2=t$ through the 
Cayley transformation.\\
The regularity condition at infinity gives the relations
$\textrm{Re}(\beta_i)=\textrm{Re}(\beta_{i\tau})=0$ and
$\textrm{Im}(\beta_i)=2-\tau\,\textrm{Im}(\beta_{i\tau})\equiv
\beta$, leaving only the parameter $\beta$ undetermined. Its value
is fixed by imposing the monodromy condition of the classical
field at $0$ and $i\tau$. The result is
\begin{equation}\label{beta}
    \beta \,=\,
-\,2\;\frac{\eta+(\,1-\eta\,)\,t^2-t^{2\,(1-2\eta)}
\left(\,1-\eta+\eta\,t^2\,\right)}{
\rule{0pt}{.4cm}t\,(\,1-t^{2\,(1-2\eta)}\,)}\;.
\end{equation}
Writing the classical field in presence of two sources of
charges $\eta$ and $\varepsilon$ as an expansion up to the first
order in $\varepsilon$, i.e.
\begin{equation}\label{varphi2 eps expansion}
    \varphi_{2}(z)\,=\,\varphi_{cl}(z)+ \epsilon\,
    \psi(z,t\hspace{.04cm})+ O(\epsilon^2\hspace{.03cm})
\end{equation}
one finds that this analysis leads to the following expression for
$\psi(z,t\hspace{.04cm})$
\begin{eqnarray}\label{psi(z,t)}
 \psi(z,t\hspace{.04cm}) \hspace{-.1cm}& = &\hspace{-.1cm} -\,
\frac{2}{w_{12}\,(y_1 \bar{y}_1-y_2\bar{y}_2)}\,
 \left\{\,
\big(\,y_1\bar{y}_1+y_2\bar{y}_2\,\big)\,
\big(\,I_{12}+\bar{I}_{12}+2\,h_0\,\big)\rule{0pt}{.4cm}\right.
\\
\rule{0pt}{.5cm} & & \hspace{5.3cm}\left.
\rule{0pt}{.4cm}-\;\bar{y}_1 y_2\,I_{11}-\,y_1\bar{y}_2\,I_{22}
 -\, y_1\bar{y}_2
\,\bar{I}_{11}-\,\bar{y}_1 y_2\,\bar{I}_{22} \,\right\}\nonumber
\end{eqnarray}
where $w_{12}=y_1y_2'-y_1'y_2=1-2\eta$ is the constant wronskian,
\begin{equation}\label{Iij(z)}
    I_{ij}(z)\,\equiv\,\int_{0}^{\,z}
    y_{i}(x)\,y_{j}(x)\,q(x)\,dx
\end{equation}
and $h_0$ is a free real parameter which cannot be determined
through monodromy arguments because it is the coefficient of a
solution of the homogeneous equation. It is fixed by requiring the
vanishing of $\psi(z,t\hspace{.04cm})$ at infinity, i.e. when
$|z|\rightarrow 1$, in order to respect the boundary condition 
(\ref{varphiB bc infinity}).
The result is
\begin{equation}\label{Reh(t)}
h_0 \,=\,
\frac{1}{2}\,\left(\,\frac{1+t^{2\,(1-2\eta)}}{\rule{0pt}{.4cm}
1-t^{2\,(1-2\eta)}}\,\log t^{2\,(1-2\eta)}+2\,\right)\;.
\end{equation}
Given the expansion (\ref{varphi2 eps expansion}) and the
Liouville equation for $\varphi_{2}(z)$, it is easy to see that
the Green function on the background $\varphi_{cl}(z)$ given by
(\ref{phiclassic}) is
$g(z,t\hspace{.04cm})=\psi(z,t\hspace{.04cm})/4$. By exploiting
the invariance under rotation, we can write our result for a
generic complex $t\in \Delta$. The final expression of the exact Green
function in the explicit symmetric form is
\begin{eqnarray}\label{g(z,t) symmetric}
 \rule{0pt}{1cm}g(\hspace{.02cm}z,t\hspace{.04cm}) & = &
 -\,\frac{1}{\rule{0pt}{.4cm}2}\;
\frac{1+(z\bar{z}\hspace{.04cm})^{1-2\eta}}
{\rule{0pt}{.4cm}1-(z\bar{z}\hspace{.04cm})^{1-2\eta}}\;
\frac{1+(\hspace{.02cm}t\bar{t}\hspace{.04cm})^{1-2\eta}}
{\rule{0pt}{.4cm}1-(\hspace{.02cm}t\bar{t}\hspace{.04cm})^{1-2\eta}}\;
\log\omega(\hspace{.02cm}z,t\hspace{.04cm})\,
-\,\frac{1}{\rule{0pt}{.4cm}1-2\eta}\,
\\
\rule{0pt}{1cm} & & \hspace{-2cm}-\,
\frac{1}{\rule{0pt}{.4cm}\,1-(z\bar{z}\hspace{.04cm})^{1-2\eta}}\;
\frac{1}{\rule{0pt}{.4cm}\,1-(\hspace{.02cm}t\bar{t}
\hspace{.04cm})^{1-2\eta}}\;\left\{\,
(z\bar{t}\hspace{.05cm})^{1-2\eta}\,\Big(\,B_{\,z/\,t}
\big(\,2\eta,\,0\,\big)-B_{\,z\bar{t}\,}\big(\,2\eta,\,0\,\big)\,\Big)
\rule{0pt}{.6cm}\right.
\nonumber \\
& &\rule{0pt}{.6cm} \hspace{4cm}\left.
+\,(\bar{z}t\hspace{.02cm})^{1-2\eta}\,\Big(\,B_{\,t/z}
\big(\,2\eta,\,0\,\big)-B_{\,1/(z\bar{t}\hspace{.02cm})\,}
\big(\,2\eta,\,0\,\big)\,\Big)
+\textrm{c.c.}\rule{0pt}{.6cm}\;\right\} \nonumber
\end{eqnarray}
where $\omega(\hspace{.02cm}z,t\hspace{.04cm})$ is the $SU(1,1)$
invariant ratio
\begin{equation}\label{SU(1,1) invariant ratio}
\omega(\hspace{.02cm}z,t\hspace{.04cm})\,=\,\frac{(z-t)\,
(\bar{z}-\bar{t}\hspace{.04cm})}
{\rule{0pt}{.4cm}\,(\hspace{.02cm}1-z\bar{t}\hspace{.04cm})\,
(\hspace{.02cm}1-\bar{z}t\hspace{.03cm})}
\end{equation}
and $B_x(a,0)$ is a particular case of the incomplete Beta
function $B_x(a,b)$ \cite{Bateman}
\begin{equation}
    B_{x}(\hspace{.02cm}a,0\hspace{.03cm})\,=\,
\frac{x^a}{a}\,F(\hspace{.02cm}a,1;a+1;\,x)\,=\,\int_0^{\,x}
    \frac{y^{a-1}}{1-y}\;dy\,=\,\sum_{n\,=\,0}^{+\infty}
\frac{x^{a+\,n}}{a+n}\;.
\end{equation}
From the expression (\ref{g(z,t) symmetric}), one can verify that
$g(z,t\hspace{.04cm})=O\big((1-z\bar{z})^2\big)$ when
$|z|\rightarrow 1$ \cite{MT: to appear}.\\
Moreover, in the limit $\eta\rightarrow 0$, we recover the
propagator $\textrm{g}(z,t\hspace{.04cm})$ on the pseudosphere
without sources given in \cite{DFJ, ZZpseudosphere}, which has the
same boundary behavior at infinity and which has been used to
perform the previous perturbative checks of the bootstrap formula
for the one point
function \cite{ZZpseudosphere, MTtetrahedron, MTgeometric}.\\

\noindent A related function playing a crucial role in the
following is the Green function at coincident points regularized
according to the ZZ procedure \cite{ZZpseudosphere}, i.e.
\begin{equation}\label{ZZ regularization}
    g(z,z)\,\equiv\,\lim_{t\,\rightarrow\,
    z}\,\left\{\,g(z,t\hspace{.06cm})+\frac{1}{2}\,\log\left|\,z-t\,
\right|^2\,\right\}\;.
\end{equation}
From (\ref{g(z,t) symmetric}), we find
\begin{eqnarray}\label{g(z,z)}
\hspace{-0cm} g(\hspace{.02cm}z,z\hspace{.03cm}) & = &
\Bigg(\,\frac{1+(z\bar{z}\hspace{.04cm})^{1-2\eta}}
{\rule{0pt}{.4cm}1-(z\bar{z}\hspace{.04cm})^{1-2\eta}}\,\Bigg)^2
\log\big(\,1-z\bar{z}\,\big)\,
-\,\frac{1}{\rule{0pt}{.4cm}1-2\eta}\;\frac{1+(z\bar{z}
\hspace{.04cm})^{1-2\eta}}{\rule{0pt}{.4cm}1-(z\bar{z}
\hspace{.04cm})^{1-2\eta}}
\\
\rule{0pt}{1cm}
 & &   +\,\frac{2\,(z\bar{z}\hspace{.04cm})^{1-2\eta}}{
\big(\rule{0pt}{.45cm}\,1-(z\bar{z}\hspace{.04cm})^{1-2\eta}\,\big)^2}\,
\left(
B_{z\bar{z}}\big(\,2\eta\,,0\,\big)+B_{z\bar{z}}\big(\,2-2\eta\,,0\,\big)
\rule{0pt}{.5cm}\right.\nonumber\\
& &\hspace{6.2cm} \left.\rule{0pt}{.5cm}+
2\gamma_E+\psi(2\eta)+\psi(2-2\eta)-\log z\bar{z}
\,\right)\nonumber
\end{eqnarray}
where $\gamma_{E}$ is the Euler constant and
$\psi(x)=\Gamma'(x)/\Gamma(x)$.
When $|z|\rightarrow 1$, $g(z,z)$ diverges logarithmically.\\

\noindent The $O(b^0)$ quantum correction to the $N$ point
function is given by the quantum determinant
$\textrm{Det}\,D\,=\,\textrm{Det}\,D(\eta_1,z_1;\dots;\eta_N,z_N)$
\begin{equation}\label{det definition}
\big(\,\textrm{Det}\,D\,\big)^{-1/2}\,\equiv\,\int\hspace{-.06cm}
\mathcal{D}\,[\, \chi \,]\;\,
\exp\left\{-\,\frac{1}{2}\,\int_\Delta \chi
\left(-\,\frac{2}{\pi}\,\partial_z\partial_{\bar{z}}\,+ \,4\mu
    b^2\,e^{\varphi_{\scriptscriptstyle\hspace{-.05cm}B}}\right)
\chi\,d^2z\,\right\}
\end{equation}
where $\varphi_{\scriptscriptstyle\hspace{-.05cm}B}$ is the
classical background field solving the Liouville equation with $N$
sources.\\
The quantum determinant can be computed by taking the logarithmic
derivative w.r.t. $\eta_j$, with $j=1,\dots,N$, and then
integrating back. The key formula is
\begin{equation}\label{det integral}
\frac{\partial}{\partial \eta_j}\,\log
\big(\,\textrm{Det}\,D\,\big)^{-1/2} =\;-\,2\mu b^2 \int_\Delta\,
g(z,z)\;\frac{\partial
e^{\varphi_{\scriptscriptstyle\hspace{-.05cm}B}}}{\partial
\eta_j}\;d^2z\hspace{1.2cm}\forall \,j\,=\,1,\dots,\, N
\end{equation}
which is a convergent integral. From this expression, the
transformation laws (\ref{varphiB transformations}) and
\begin{equation}\label{g(z,z) SU(1,1) transformation }
    g(z,z)
    \;\;\longrightarrow\;\;\tilde{g}(w,w)\,=\,g(z,z) +
    \,\frac{1}{2}\,\log \left|\,\frac{dw}{dz}\,\right|^2
\end{equation}
we can see that the semiclassical dimensions of the vertex
operators $V_\alpha(z)$ get corrected to $\Delta_\alpha =
\alpha(1/b+b-\alpha)$, which are the quantum dimensions obtained
from the hamiltonian approach on the sphere \cite{CT}. It is also
possible to prove that higher orders corrections do not change
these dimensions \cite{MT: to
appear}.\\

\noindent In the case of the one point function, the expression
(\ref{det integral}) can be explicitly computed and the result is
\begin{equation}\label{log det integral}
\frac{\partial}{\partial \eta}\,\log
\big(\,\textrm{Det}\,D(\eta,0)\,\big)^{-1/2}
=\;2\,\gamma_{\,\scriptscriptstyle\hspace{-.05cm}E}\, +\,2\,
\psi(1-2\eta)\,+\,\frac{3}{1-2\eta}\;.
\end{equation}
Integrating back in $\eta$ with the initial condition given in
\cite{ZZpseudosphere}, i.e. $
\left.\big(\,\textrm{Det}\,D(\eta,0)\,\big)^{-1/2}\,\right|_{\,\eta\,=\,0}
\hspace{-.2cm}=\,1$, we find
\begin{equation}\label{quantum det eta}
\log \big(\,\textrm{Det}\,D(\eta,0)\,\big)^{-1/2}\,=\,
2\,\gamma_{\,\scriptscriptstyle\hspace{-.05cm}E} \,\eta-\log
\Gamma(1-2\eta) -\frac{3}{2}\,\log(1-2\eta)\;.
\end{equation}
Putting this result together with the classical contribution
(\ref{one point classical term}), we have
\begin{eqnarray}\label{one point serie b}
\rule{0pt}{.6cm} \log \left\langle \, V_{\eta/b}(0)
\,\right\rangle  \hspace{-.1cm} & = & \hspace{-.2cm}
-\,\frac{1}{b^2}\;\Big(\,\eta\,\log\left[\,\pi
        b^2\mu\,\right]+2\eta+(1-2\eta)\,\log(1-2\eta)\,\Big)
        \nonumber\\
 \rule{0pt}{.8cm} & & +\,\Big(\,
2\,\gamma_{\,\scriptscriptstyle\hspace{-.05cm}E} \,\eta-\log
\Gamma(1-2\eta) -\frac{3}{2}\,\log(1-2\eta)\,\Big) + O(b^2)
\end{eqnarray}
to all orders in $\eta$.\\
We can compare (\ref{one point serie b}) with the result obtained
by ZZ within the bootstrap approach \cite{ZZpseudosphere}.
Conformal invariance imposes the following structure for the one
point function
\begin{equation}
 \label{one-point function}
\left\langle \,V_{\alpha}(z_1) \,\right\rangle =
\frac{U(\alpha)}{(\,1-z_1\bar{z}_1\,)^{\,2\alpha(Q-\alpha)}}
\end{equation}
where $U(\alpha)$ is the one point structure constant. $U(\alpha)$
has been determined through the bootstrap method
\cite{ZZpseudosphere} and the result for the basic vacuum is
\begin{equation}
U(\alpha)\,=\,U_{1,1}(\alpha)\,=\,\big(\,\pi\mu\gamma(b^2)\,\big)^{-\alpha/b}
\frac{\Gamma(Q\,b)\,\Gamma(Q/b)\,Q}{\rule{0pt}{.36cm}
\Gamma\big((Q-2\alpha)\,b\big)\,\Gamma\big((Q-2\alpha)/b\big)\,(Q-2\alpha)}
\end{equation}
where $\gamma(x)=\Gamma(x)/\,\Gamma(1-x)$.\\
Our result (\ref{one point serie b}) agrees with the expansion in
$b^2$ of $U(\eta/b)$. We stress that (\ref{one point serie b})
corresponds to the summation of an infinite class of graphs of the
usual perturbative expansion 
\cite{ZZpseudosphere, MTtetrahedron, MTgeometric}.\\

\noindent The technique developed above can be applied to compute
the two point function with one arbitrary heavy charge $\eta$ and
another heavy charge $\varepsilon$
\begin{equation}\label{2 point vertex eta-epsilon}
\left\langle \, V_{\eta/b}(0)\,
V_{\varepsilon/b}(t)\,\right\rangle
\end{equation}
up to the order $O(\varepsilon)$  and $O(b^0)$ included, but to
all orders in $\eta$ and $t$.\\
First we observe that it is more convenient to work with the
following ratio
\begin{equation}\label{conformal block}
  g_{\,\eta/b,\,\varepsilon/b}\big(\,\omega(0,t)\,\big)\,\equiv\, 
\frac{\left\langle \, V_{\eta/b}(0)\,
V_{\varepsilon/b}(t)\,\right\rangle}{\rule{0pt}{.41cm}\left\langle
\, V_{\eta/b}(0)\,\right\rangle \left\langle \,
 V_{\varepsilon/b}(t)\,\right\rangle}
\end{equation}
where $\omega(z,t)$ is the $SU(1,1)$ invariant ratio (\ref{SU(1,1)
invariant ratio}). Then, the two point function up to the orders
$O(\varepsilon)$ and $O(b^0)$ is given by
\begin{eqnarray}\label{2 point b serie }
\left\langle \, V_{\eta/b}(0)\,V_{\varepsilon/b}(t)\,\right\rangle
&  = &
e^{-S_{cl}(\eta,\,0;\,\varepsilon,\,t)}\,
\big(\,\textrm{Det}\,D(\eta,\,0)\,\big)^{-1/2}\,\times\\
    &  &
\rule{0pt}{.8cm} \hspace{-.15cm} \times\,\left(\,1\, -\,8\,\mu
b^2\varepsilon \int_{\Delta}g(z,t)\,e^{\varphi_{cl}(z)}
g(z,z)\,d^2z\,+O(\varepsilon^2)\,\right)\Big(\,1+O(b^2)\,\Big)
\nonumber
\end{eqnarray}
where $S_{cl}(\eta,0;\varepsilon,t)$ is the classical action
(\ref{action classical phiB}) with $N=2$ evaluated on the
classical field $\varphi_{2}(z)$, given in (\ref{varphi2 eps
expansion}), which describes the pseudosphere with a curvature
singularity of charge $\eta$ in $z_1=0$ and another curvature
singularity of charge
$\varepsilon$ in $z_2=t$, up to $O(\varepsilon^2)$.\\
For the logarithm of the ratio (\ref{conformal block}), we find
\begin{eqnarray}\label{2 point log integrals}
\log\,\frac{\left\langle \, V_{\eta/b}(0)\,
V_{\varepsilon/b}(t)\,\right\rangle}{\rule{0pt}{.41cm}\left\langle
\, V_{\eta/b}(0)\,\right\rangle \left\langle \,
 V_{\varepsilon/b}(t)\,\right\rangle}
 & = & \frac{\varepsilon}{b^2}\;
 \Big\{\,\varphi_{cl}(t)-\left.\varphi_{cl}(t)\right|_{\,\eta\,=\;0}\Big\}\\
 \rule{0pt}{.8cm}& &\hspace{-2.4cm}
- \,8\,\mu b^2\varepsilon \,\left\{\,
\int_{\Delta}\hspace{-.07cm}g(z,t)\,e^{\varphi_{cl}(z)}
g(z,z)\,d^2z\, -
\int_{\Delta}\hspace{-.07cm}\textrm{g}(z,t)\,
e^{\left.\varphi_{cl}(t)\right|_{\,\eta\,=\;0}}
\,\textrm{g}(z,z)\,d^2z\,\right\}\;.\nonumber
\end{eqnarray}
It is possible to show that the r.h.s. of (\ref{2 point log
integrals}) is invariant under $SU(1,1)$ transformations, as
expected, and both the integrals occurring in (\ref{2 point log
integrals}) can be explicitly computed. The final result is
\begin{eqnarray}\label{2 point final result}
\log\,\frac{\left\langle \, V_{\eta/b}(z)\,
V_{\varepsilon/b}(t)\,\right\rangle}{\rule{0pt}{.41cm}\left\langle
\, V_{\eta/b}(z)\,\right\rangle \left\langle \,
 V_{\varepsilon/b}(t)\,\right\rangle}
 & = &
 \frac{\varepsilon}{b^2}\;
 \left\{-\log\frac{\big(\,\omega^\eta-\omega^{1-\eta}\,\big)^2}
 {\rule{0pt}{.38cm}(1-2\eta)^2}
 \,+\,\log(\hspace{.02cm}1-\omega)^2\,\right\}\nonumber\\
\rule{0pt}{1cm} & & \hspace{-4.8cm}+\;\varepsilon\,
 \left\{\,\frac{2}{\rule{0pt}{.4cm}\big(1-\omega^{1-2\eta}\big)^2}\,
 \left(\,B_{\omega}(2-2\eta,0)+\psi(2-2\eta)+
\gamma_{\,\scriptscriptstyle\hspace{-.05cm}E}
         +\frac{1}{2(1-2\eta)}\rule{0pt}{.7cm}\right.\right.
         \\
\rule{0pt}{.7cm} & & \hspace{-1.5cm} +\;\omega^{2(1-2\eta)}
\left(\,B_{\omega}(2\eta,0)+\psi(2\eta)+
\gamma_{\,\scriptscriptstyle\hspace{-.05cm}E}+\frac{3}{2(1-2\eta)}\,
-\log\omega\,\right)       \nonumber\\
\rule{0pt}{.9cm} & &\hspace{-.7cm}\left. \rule{0pt}{.7cm}
\left.+\,2\,\omega^{1-2\eta}\left(\,\log(1-\omega)
-\,\frac{1}{1-2\eta}\;\right)\rule{0pt}{.7cm}\right)
+\,2\,\log(1-\omega)\,-\,3\;\right\}\;\nonumber
\end{eqnarray}
where $\omega=\omega(z,t)$ is given by (\ref{SU(1,1) invariant
ratio}).\\
This expression satisfy the cluster property
\cite{ZZpseudosphere}, i.e. $g_{\,\eta/b,\,\varepsilon/b}
(\omega)\rightarrow 1$ when $\omega\rightarrow 1$.\\
Setting $\alpha_1=\eta/b$ and $\alpha_2=\varepsilon/b$, we have
checked that (\ref{2 point final result}) agrees with the standard
perturbative expansion up to $O(\alpha_1\alpha_2\,b^2)$ included
\cite{MTgeometric}. A second check consists in the comparison of
(\ref{2 point final result}) with the degenerate two point
function with $\alpha_1=-1/(2b)$, i.e. $\eta=-1/2$, and
$\alpha_2=\varepsilon/b$. In this case the ratio (\ref{conformal
block}) is explicitly known \cite{ZZpseudosphere, FZZ,
difrancesco}
\begin{equation}\label{degenerate 2 point}
g_{\,-1/(2b),\,\varepsilon/b}(\omega)\,=\,
\omega^{\varepsilon/b^2}\, _2
F_1\big(\,1+1/b^2,\,2\,\varepsilon/b^2;\,2+2/b^2;\,1-\omega\,\big)
\end{equation}
and we find that the expansion of the logarithm of
(\ref{degenerate 2 point}) up to $O(\varepsilon)$ and $O(b^0)$
included reproduces (\ref{2 point
final result}) computed for $\eta=-1/2$.\\
In \cite{ZZpseudosphere}, ZZ gave the ratio (\ref{conformal
block}) in terms of a conformal block with null intermediate
dimension as follows
\begin{equation}\label{conformal block 2}
    g_{\,\alpha_1,\,\alpha_2}(\omega)\,=\,
(1-\omega)^{2\Delta_{\hspace{.03cm}\alpha_{1}}}
 \,\mathcal{F}\left(
 \begin{array}{cc}
\alpha_1 & \alpha_2\\
\alpha_1 & \alpha_2
 \end{array};\,i\,Q/2,\,1-\omega \right)\;.
\end{equation}
According to this statement, (\ref{2 point final result}) provides
an expansion up to $O(\varepsilon)$ and $O(b^0)$, but to all
orders in $\eta$ and $\omega$, of the conformal block
(\ref{conformal block 2}) with $\alpha_1=\eta/b$ and 
$\alpha_2=\varepsilon/b$.\\
An extension of the described technique to boundary Liouville
field theory \cite{FZZ} is underway \cite{MT: to appear}.\\


\end{document}